\documentclass[fleqn,usenatbib]{mnras}

\usepackage[T1]{fontenc}

\DeclareRobustCommand{\VAN}[3]{#2}
\let\VANthebibliography\thebibliography
\def\thebibliography{\DeclareRobustCommand{\VAN}[3]{##3}\VANthebibliography}

\usepackage{graphicx}	
\usepackage{amsmath}	
\usepackage{amssymb}	
\usepackage{multirow}
\usepackage[czech]{babel}
\usepackage{newtxtext,newtxmath}

\title[NGC~3147]{NGC~3147: a prototypical low-luminosity active galactic nucleus with double-peaked optical and ultraviolet lines}

\author[Stefano Bianchi et al.]{
Stefano Bianchi,$^{1}$\thanks{E-mail: bianchi@fis.uniroma3.it (SB)}, Marco Chiaberge$^{2,3}$, Ari Laor$^4$, Robert Antonucci$^5$, Atharva Bagul$^6$, Alessandro Capetti$^7$
\\
$^{1}$Dipartimento di Matematica e Fisica, Universit\`a degli Studi Roma Tre, via della Vasca Navale 84, 00146 Roma, Italy\\
$^2$Space Telescope Science Institute, 3700 San Martin Dr., Baltimore, MD 21210, USA\\
$^3$Johns Hopkins University, 3400 N. Charles Street, Baltimore, MD 21218, USA\\
$^4$Physics Department, Technion - Israel Institute of Technology, Haifa 32000, Israel\\
$^5$Department of Physics, University of California, Santa Barbara, CA, USA\\
$^6$Department of Physics, Indian Institute of Science Education and Research, Bhopal, Madhya Pradesh 462066, India\\
$^7$INAF - Osservatorio Astrofisico di Torino, Via Osservatorio 20, I-10025 Pino Torinese, Italy\\
}

\date{Accepted XXX. Received YYY; in original form ZZZ}

\pubyear{2021}

\begin{document}
\label{firstpage}
\pagerange{\pageref{firstpage}--\pageref{lastpage}}
\maketitle

\begin{abstract}
A previous narrow-slit ($0.1$ arcsec) \textit{Hubble Space Telescope} observation unveiled a broad relativistic H$\alpha$ profile in NGC~3147, a low-luminosity ($\mathrm{L_{bol}}\sim10^{42}$ erg s$^{-1}$), low-Eddington ratio ($\mathrm{L_{bol}/L_{Edd}}\sim10^{-4}$) active galactic nucleus (AGN), formerly believed to be a candidate true type 2 AGN intrinsically lacking the broad-line region. The new observations presented here confirm the double-peaked profile of the H$\alpha$ line, which further shows variability both in flux and in the inner radius of the emitting disc with respect to the previous epoch. Similar disc line profiles are also found in prominent ultraviolet (UV) lines, in particular Ly$\alpha$ and \ion{C}{iv}. The new data also allow us to build a simultaneous subarcsec optical-to-X-ray spectral energy distribution of NGC~3147, which is characterized by the absence of a thermal UV bump, and an emission peak in the X-rays. The resulting very flat $\alpha_{ox}=-0.82$ is typical of low-luminosity AGN, and is in good agreement with the extrapolation to low luminosities of the well-known trend with luminosity observed in standard AGN. Indeed, we are possibly observing the accretion disc emission in NGC~3147 in the optical, close to the expected peak. On the other hand, the steep -2 UV power law may be Comptonization of that cold disc by a warm corona, what is instead generally observed as a `soft excess' in more luminous AGN. 

\end{abstract}




\section{Introduction}

NGC~3147 is a low-luminosity ($\mathrm{L_{bol}}\sim10^{42}$ erg s$^{-1}$), low-accreting ($\mathrm{L_{bol}/L_{Edd}}\sim10^{-4}$) active galactic nucleus (AGN), considered as the best case of true type 2 Seyfert \citep{Panessa2002,Bianchi2008c,Bianchi2012,Bianchi2017a,Matt2012}, thanks to simultaneous optical and X-ray observations, which unambiguously verified an absorption-free line of sight to the nucleus together with no evidence for broad optical lines. 
However, the latter point was critically dependent on contamination from the host emission. Indeed, the \textit{Hubble Space Telescope} (HST) Space Telescope Imaging Spectrograph (STIS) G750L spectrum of NGC~3147 taken in 2018, thanks to its narrow slit ($0.1$ arcsec), allowed for a significant reduction of the host galaxy emission with respect to ground-based spectra, revealing an otherwise hidden very broad (full width at zero intensity of $\sim27\,000$ km s$^{-1}$) H$\alpha$ emission line \citep[][]{Bianchi2019a}. 

The luminosity of the broad H$\alpha$ line in the HST spectrum ($1.3\times10^{40}$ erg s$^{-1}$) is in agreement with the expectations from scaling relations in type 1 AGN based on X-ray and [\ion{O}{iii}] emission  \citep{Stern2012,Stern2012c}, extending them at the very low end of luminosity range. Even more surprisingly, the very broad and asymmetric profile of the H$\alpha$ line is characterized by a steep cutoff blue wing and an extended red wing, i.e. the clear signature of emission from a mildly relativistic thin accretion disc line \citep[e.g.][]{Laor1991}. The profile is accurately modelled with the expected emission from a disc with an inclination of $\sim23^{\circ}$, extending from $\sim77$ to $\sim570$ r$_g$, where r$_g=GM/c^2$ is the gravitational radius \citep[][]{Bianchi2019a}. The inner radius derived from the disc line profile fit agrees well with the distance to the broad-line region (BLR) as derived from the standard $R_\mathrm{BLR}$ vs. $L_{5100}$ relation \citep[e.g.][]{Greene2005}, again extending it down to a much lower luminosity range.

The implications of these results are far-reaching for multiple reasons. First of all, NGC~3147 is definitely not a true type 2 AGN, suggesting that the entire class of objects may not exist at all \citep[see also][for a similar conclusion]{Antonucci2012}. A low-accreting, high-mass, object is expected to have a very small radius of the BLR, both in absolute size, and in r$_g$ for a given black hole (BH) mass. The resulting very broadened line profile, combined with the intrinsic weakness with respect to the host galaxy, makes it difficult to disentangle the BLR emission, unless the host contamination is significantly reduced thanks to narrow-slit observations. 

Very broad, H$\alpha$ disc line profiles have been previously observed in few other low-luminosity AGN (LLAGN) \citep[e.g.][]{Eracleous1994,Eracleous2003,Tran2010a, Balmaverde2014b,Storchi-Bergmann2016}, typically with inner radii in the range $10^2-10^3 r_g$. The very small radius measured for NGC~3147 is suggestive of the outer part of the thin accretion disc  which powers higher luminosity AGN. Its very presence is surprising at such low accretion rates, whereby instead an optically thin and quasi-spherical accretion flow is commonly assumed \citep[e.g.][]{Blandford1999}. Moreover, the standard scaling relations for the luminosity and the radius of the BLR have been found to hold also at these very low luminosity and accreting regimes. These results seem to indicate that LLAGN may be more similar to luminous AGN than previously thought.

In this paper, we report on new \textit{HST} observations of NGC~3147. The main objective is to confirm the properties of the disc line profile of the H$\alpha$ line, by taking advantage of a higher resolution spectrum, and reveal any possible variability of its flux and profile parameters. Moreover, we extend the analysis to the ultraviolet (UV) spectrum, looking for the same disc line profiles in other prominent emission lines as Ly$\alpha$ and \ion{C}{iv}. Finally, we build the first simultaneous subarcsec UV-to-X-ray spectral energy distribution (SED) of an LLAGN, and compare it to other LLAGN and standard luminous AGN. 

\section{Observations and data reduction}\label{datared}

The target was observed with the STIS instrument on board the {\it HST}, as part of programme GO-15908 (PI: S. Bianchi). Observations executed on 2020 March 7. The data were taken over four orbits, using both the CCD and the far-ultraviolet (FUV) Multi-Anode Microchannel Array (MAMA) detectors. The first orbit was used to observe NGC~3147 with the G750M grism. Two exposures were taken to allow for cosmic ray and detector defect rejection. The total exposure time for G750M was 2235 s. In the second orbit we used the G750L and 2x400s to obtain a spectrum for comparison with the previous observations taken in 2018. The remaining part of the second orbit and orbits 3 and 4 were used with the FUV/MAMA and G140L, for a total exposure time of 8379s. The 52x0.1 slit was used for all observations. A flat field was also taken with the CCD to better remove fringes, as recommended by the STIS Instrument Handbook \citep[][]{Prichard2022}.

All files were downloaded from the Mikulski Archive for Space Telescopes (MAST). For the CCD data, we (flux and wavelength) calibrated all files using the python pipeline that includes correction of charge transfer efficiency losses \citep[][]{Anderson2010}. At this stage we also applied the flat field for G750L obtained during our observations. The calibrated STIS/MAMA observations were downloaded from the MAST and no post-processing recalibration was needed.

For G750L and G750M we extract the spectra from a 3-pixel aperture, corresponding to a projected angular size of 0.15 arcsec. In addition, we re-extracted the spectra taken as part of the Cycle 25 programme (GO-15225, PI S. Bianchi), published in \citet{Bianchi2019a}, using the same aperture to allow for a more accurate comparison of the two epochs. The flux calibration is performed as described in the STIS Data Handbook \citep[][]{Sohn2019} assuming that the source is unresolved. We thus multiply the values derived from the extraction by the {\it DIFF2PT} keyword and then apply the appropriate wavelength-dependent corrections for the used non-standard extraction slit height. For the co-added spectra taken with the MAMA we use the default extraction height of 11 pixels, corresponding to 0.275 arcsec projected on the sky. We tested that the only source is the nucleus at these wavelengths, therefore a smaller extraction height would only lead to a reduced signal-to-noise ratio (S/N). For the extraction of all spectra we used the \textsc{pyraf} task {\it splot}. Errors are extracted from the error array for each data set, and added in quadrature. For the CCD, we divide the error by the square root of the extraction height, in pixels. This should not be done for the MAMA, because of the different nature of the detector. 

We can give an estimate of the starlight contribution in the observed optical red continuum of our G750L spectrum. The image has a spike consistent with a point source, and it is clearly separable from the diffuse starlight. We found a star image from another programme with the same optical elements and detector, to use for a model point spread function (PSF; star G158-100, Proposal ID: 7642). Its profile matches our spike in NGC~3147 very well. We performed a series of PSF subtractions on our image using increasing normalization until the residual showed a flat top. That provides an upper limit to the point source. We then made a small correction for the emission line contribution to the point source, and concluded that the point source fraction in the optical red G750L spectrum in our spectroscopic aperture is $0.24\pm 0.05$. Note that we allowed the starlight to be flat-topped, while it is likely to be peaked. That is a small effect however. Further details on our estimate of the host contamination are provided in Appendix~\ref{host}.

\textit{Swift}'s X-ray Telescope (XRT) observed NGC~3147 on 2020 March 7 for $\sim1.6$ ks (obsid 00081688003), simultaneously with the \textit{HST} G750M pointing, as a Target of Opportunity triggered by the \textit{HST} observation. Data were automatically reduced and the XRT spectrum extracted in a circular region with a radius of 20 pixels ($\sim47$ arcsec) via the interactive tools available at the ASI Space Science Data Center\footnote{\url{https://www.ssdc.asi.it/}}.

Spectra were analysed with \textsc{xspec} 12.12.0 \citep{Arnaud1996}, and the fitting model convolved with a Gaussian smoothing with $\sigma=3.14$ \AA\ (G750L), $\sigma=0.358$ \AA\ (G750M), and $\sigma=0.572$ \AA\ (G140L) to match the corresponding instrumental resolutions. All the reported velocity shifts are relative to the adopted systemic redshift of the host galaxy \citep[z=0.009346:][]{Epinat2008}. In the following, (statistical only) errors and upper limits correspond to the 90 per cent confidence level for one interesting parameter, apart from the plots, where $1\sigma$ error bars are shown. The adopted cosmological parameters are $H_0 = 70$ km s$^{-1}$ Mpc$^{-1}$, $\Omega_\Lambda = 0.73$ and $\Omega_m = 0.27$.

\section{Spectral Analysis}

\subsection{\label{optan}The optical spectrum}

\begin{figure}
	\includegraphics[width=\columnwidth]{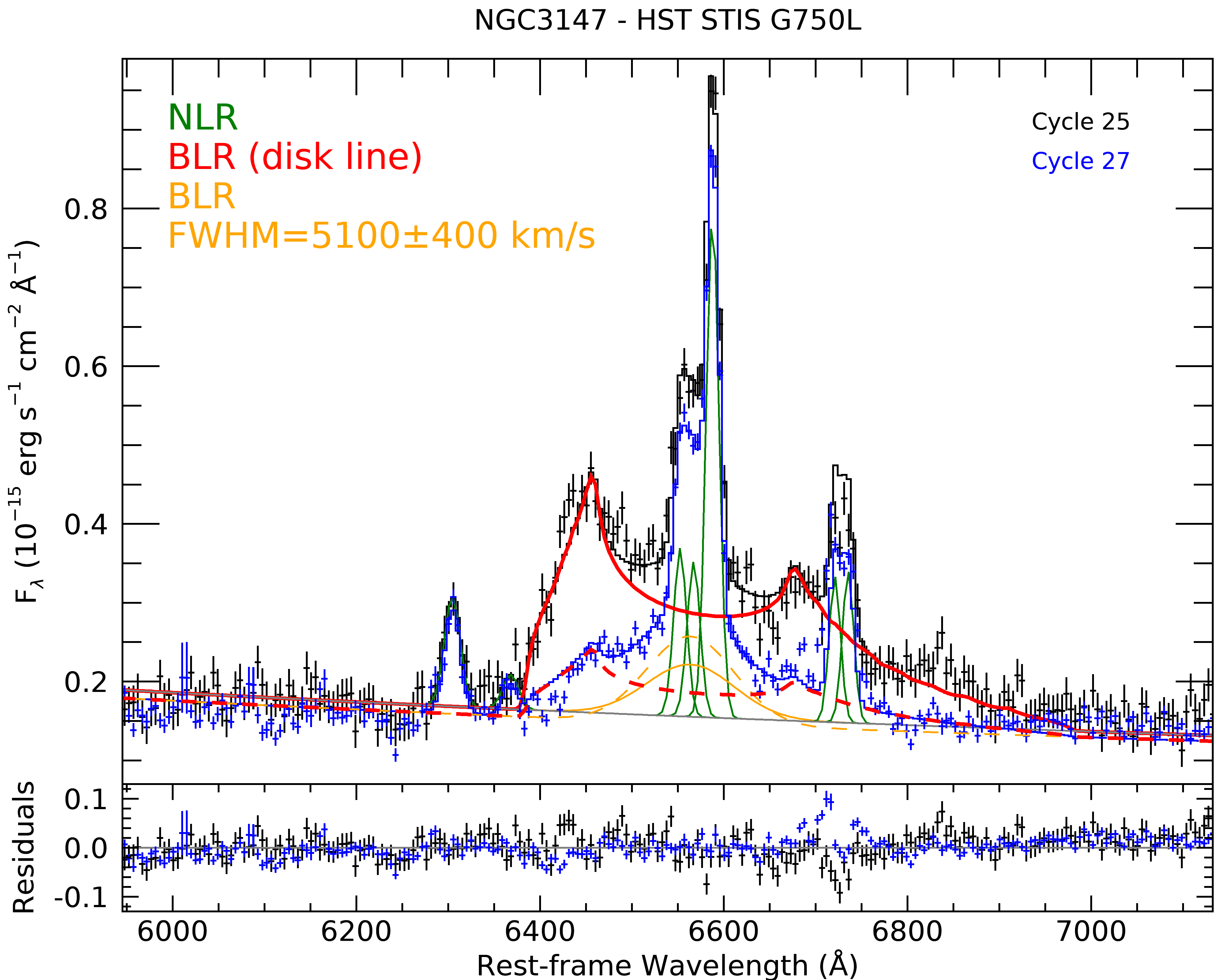}
    \caption{A comparison between the HST STIS G750L spectrum of NGC 3147 taken in Cycle 25 (black) and the new one taken in Cycle 27 (blue). The broad relativistic line has a significantly lower flux in the new data (dashed red), allowing the detection of a further broad Gaussian component for the H$\alpha$, mildly variable between the two cycles (solid and dashed orange). Narrow lines from the NLR are in green and the continuum in gray. The fit residuals are shown in the bottom panel.}
    \label{fig:kerrdisk}
\end{figure}

We start our analysis by comparing the new G750L spectrum with the one taken in Cycle 25 and published in \citet{Bianchi2019a}. As evident from Fig.~\ref{fig:kerrdisk}, the H$\alpha$ profile changed dramatically between the two observations. Most of the variation can be accounted for a lower flux of the relativistic component by a factor of $3.2\pm0.2$ in the new data, with all the other parameters being the same. This allows for the detection of a further broad Gaussian component of the H$\alpha$ emission line, with a full width at half-maximum (FWHM) $\sim5000$ km s$^{-1}$. This component is consistent with being present also in the old data, with a flux $1.8\pm0.3$ times lower, but was not required by the fit because of the dominance of the relativistic line. All the narrow emission lines from the narrow-line region (NLR) are consistent with being constant, together with the continuum level. If a cross-normalization factor is included between the two data sets, an $\sim5\%$ excess is found in the new data, well within any calibration uncertainties.

This best=fitting model gives a very good representation of both data sets, with the most significant residuals being bluewards the [\ion{S}{II}] doublet. We will investigate this issue with more detail below, with the high-resolution G750M spectrum. All the other parameters of the \textsc{kerrdisk} model \citep[see][for details]{Bianchi2019a} do not change significantly if allowed to be different between the two cycles, with the significant exception of the inner radius of the disc, which changed from $77\pm15$ to $195\pm15$ r$_g$ (see Table~\ref{ngc3147_lines}). The underlying continuum is well modelled by a power law in the 5450-8000 \AA\ band\footnote{The remaining 8000-9999 \AA\ band is not used for the fit, since the wavelength-dependent corrections used for the non-standard extraction slit height (see Sect.~\ref{datared}) are significantly different with respect to lower wavelengths.}, with an index (in $F_\nu$) of $-1.40\pm0.04$, and a flux density of $F_\lambda=1.44\pm0.01\times10^{-16}$ erg cm$^{-2}$ s$^{-1}$ \AA$^{-1}$ at 7000 \AA\footnote{As reported in Sect.~\ref{datared} and Appendix~\ref{host}, the red part of the G750L spectrum is likely contaminated by significant starlight, so that the intrinsic nuclear flux in this band is lower and the slope flatter than what observed.}. Galactic reddening is taken into account with an extinction of $E(B-V)=0.022$, as derived from NED based on the prescription of \citet{Schlafly2011} and variation with wavelength as specified by \citet{Cardelli1989}.

\begin{figure}
	\includegraphics[width=\columnwidth]{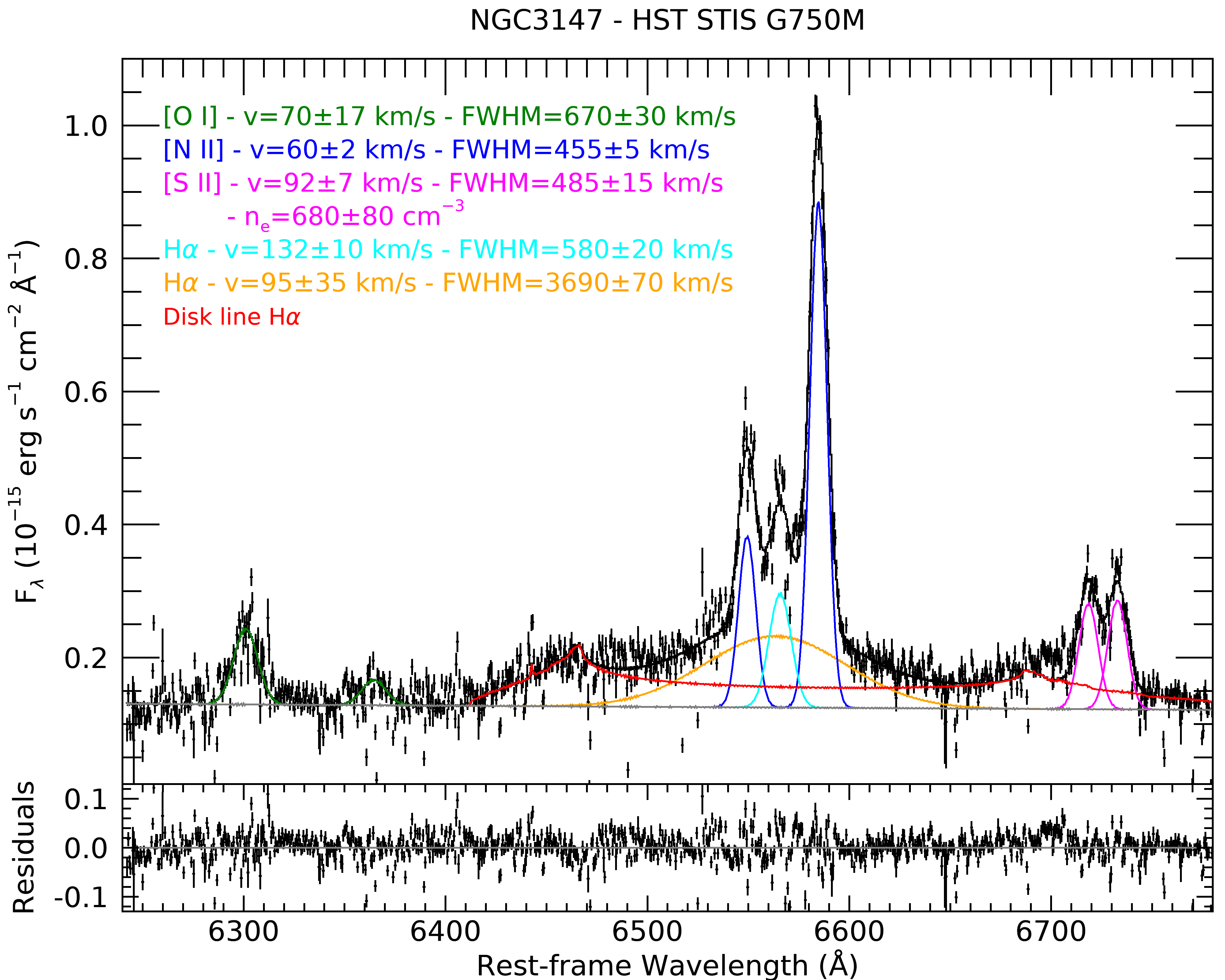}
    \caption{The HST STIS G750M spectrum of NGC~3147. As in Fig.~\ref{fig:kerrdisk}, the broad relativistic line is in red, the broad Gaussian H$\alpha$ component is in orange, and the continuum in gray. The different narrow emission lines are in other colours. The fit residuals are shown in the bottom panel.}
    \label{fig:G750M}
\end{figure}

The overall model used for G750L is qualitatively a good representation for the higher resolution G750M spectrum, when a cross-normalization is taken into account to correct for the extended nature of the source. A very good fit is found just by allowing the parameters to readjust (Fig.~\ref{fig:G750M}). While the parameters for the relativistic line remain unchanged, the FWHM of the further broad Gaussian component for the H$\alpha$ is smaller and much better constrained ($3760\pm80$ km s$^{1}$). Note that its velocity shift is comparable to that of the narrow emission lines. All the best-fitting parameters relative to the narrow emission lines are reported in Table~\ref{ngc3147_lines}. 

In particular, for the doublets we calculated the ratio of the two components. While we recover the expected ratio ($R=0.34\pm0.01$) for the [\ion{N}{II}] doublet, the ratio for the [\ion{O}{I}] is fixed to the expected value of 3. On the other hand, the observed [\ion{S}{II}] doublet ratio ($R=0.92\pm0.03$) allows us to estimate the density of the gas to $n_e=680\pm80$ cm$^{-3}$ \citep{Sanders2016}.

\begin{table}
\caption{\label{ngc3147_lines}Emission line properties in the G750M and G750L (H$\alpha$ disc line) spectra. Laboratory wavelengths (\AA) in air are from \citet{Bowen1960}. FWHMs and shift velocities are in km s$^{-1}$, radii in r$_g$ and dereddened fluxes in $10^{-15}$ erg cm$^{-2}$ s$^{-1}$. The ratio for the [\ion{O}{I}] doublet is fixed to 3 and labelled with a *.}
\begin{center}
\begin{tabular}{llllll}
\hline
Line & $\lambda_\mathrm{l}$ & FWHM & Flux & $v$ & $R$ \\
\hline
$\mathrm{[{O\,\textsc{i}}]}$ & 6300.32 & \multirow{2}{*}{$670\pm30$} & $1.83\pm0.04$ & \multirow{2}{*}{$70\pm17$} & \multirow{2}{*}{$3^*$}\\[1ex]
$\mathrm{[{O\,\textsc{i}}]}$ & 6363.81 &  & $0.61\pm0.02$& &\\[1ex]
$\mathrm{[{N\,\textsc{ii}}]}$ & 6548.06 & \multirow{2}{*}{$455\pm5$} & $3.00\pm0.07$& \multirow{2}{*}{$60\pm2$}& \multirow{2}{*}{$0.34\pm0.01$} \\[1ex]
$\mathrm{[{N\,\textsc{ii}}]}$ & 6583.39  & & $8.77\pm0.07$& &\\[1ex]
H$\alpha$ & 6562.79 & $580\pm20$ & $2.60\pm0.07$ & $132\pm10$ & --\\[1ex]
H$\alpha$ & 6562.79 & $3760\pm80$ & $9.9\pm0.2$ & $30\pm40$ & --\\[1ex]
$\mathrm{[{S\,\textsc{ii}}]}$ & 6716.42 & \multirow{2}{*}{$485\pm15$} & $1.95\pm0.06$& \multirow{2}{*}{$92\pm7$} & \multirow{2}{*}{$0.92\pm0.03$}\\[1ex]
$\mathrm{[{S\,\textsc{ii}}]}$ & 6730.78 & & $2.13\pm0.06$& &\\
\hline
\multicolumn{5}{c}{\textsc{Broad disc line profiles}}\\
$\lambda_\mathrm{l}$ & $i$ & r$_{\rm in}$ & r$_{\rm out}$ & Flux \\
\hline
\multicolumn{5}{c}{\textsc{H$\alpha$}}\\
6562.79 & $23\pm2$ & $195\pm15$ & $650\pm40$ & $16.5\pm0.6$ \\
\end{tabular}
\end{center}
\end{table}

As already noted for the G750L, positive residuals are present bluewards the [\ion{S}{II}] doublet, approximately at the location of the peak of the red wing of the relativistic H$\alpha$  line. 
Since these residuals occur close to the disc line peak velocity, they may indicate that the thin flat Keplerian disc model, with a sharp inner boundary, may be an oversimplification of the inner disc boundary.

\subsection{\label{uvan}The UV spectrum}

\begin{figure*}
	\includegraphics[width=\textwidth]{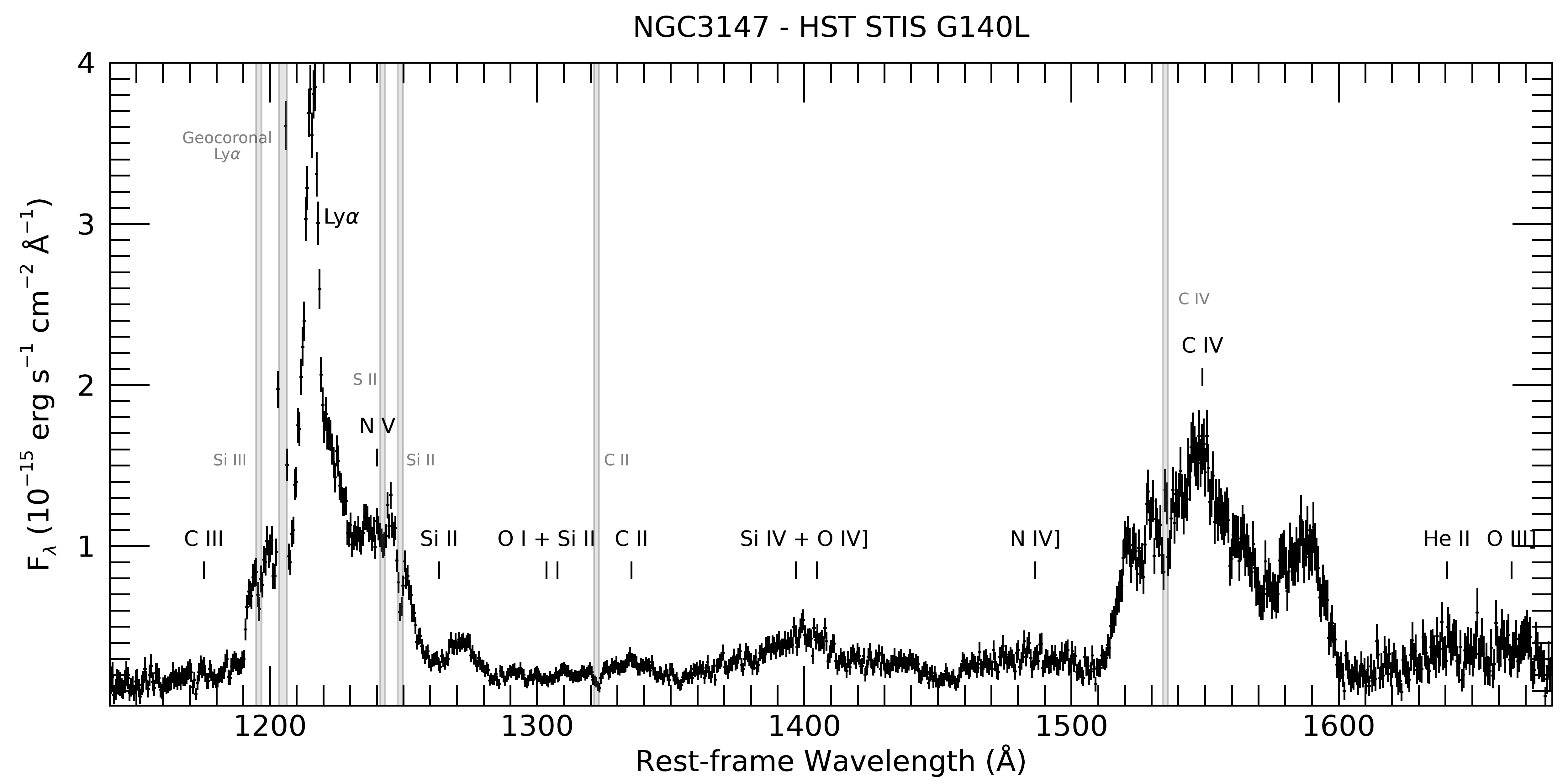}
    \caption{The \textit{HST} STIS G140L spectrum of NGC~3147. The most important expected emission lines are labelled. The geocoronal Ly$\alpha$ emission line and the foreground ISM absorption lines are instead highlighted by grey bands.}
    \label{fig:G140L}
\end{figure*}

The \textit{HST} STIS G140L spectrum is shown in Fig.~\ref{fig:G140L}. It is dominated by prominent broad Ly$\alpha$ and \ion{C}{IV} emission lines, with complex profiles. Several foreground absorption lines due to our own interstellar medium (ISM) are apparent. They are modelled as Gaussian lines when appropriate in the following fits. Once corrected for the Galactic reddening, the underlying continuum is well modelled by a power law in the whole 1150-1680 \AA\ band, with an index (in $F_\nu$) of $-2.0^{+0.3}_{-0.4}$, and a flux density of $F_\lambda=1.90\pm0.05\times10^{-16}$ erg cm$^{-2}$ s$^{-1}$ \AA$^{-1}$ at 1300 \AA.

\begin{figure}
	\includegraphics[width=\columnwidth]{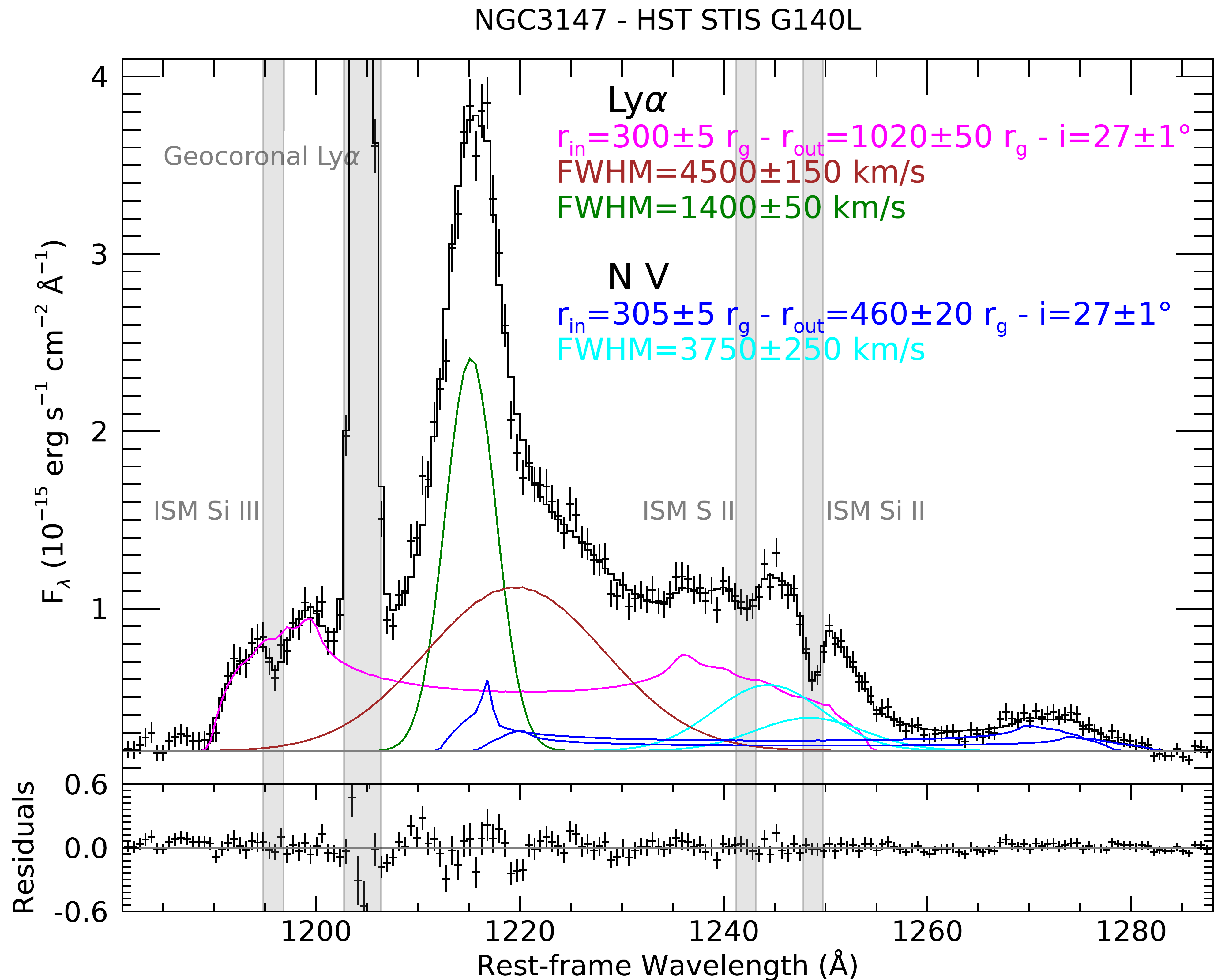}
    \caption{The Ly$\alpha$ emission line in the HST STIS G140L spectrum of NGC~3147. A detailed decomposition of the best-fitting model is shown, together with the main parameters (see text for details). The fit residuals are shown in the bottom panel. The geocoronal Ly$\alpha$ emission line and the foreground ISM absorption lines are instead highlighted by grey bands.}
    \label{fig:Lya}
\end{figure}

The Ly$\alpha$ profile can be modelled with three components, as for the H$\alpha$ line: a disc line (\textsc{Kerrdisk} model), a broad and a narrow Gaussian line (see Fig.~\ref{fig:Lya} and Table~\ref{ngc3147_G140L_lines}). The disc line profile is characterized by inner and outer radii r$_{in}=300\pm5$ and r$_{out}=1020\pm50$ r$_g$, and an inclination $i=27\pm1\degr$. The inclination angle is in very good agreement with that found for H$\alpha$, and the locations of the blue and red peaks are again just where predicted by a simple disc emission. Both the inner and the outer radii are larger than that required by the H$\alpha$ profile.

The overall Ly$\alpha$ emission line is completed by a broad (FWHM$\sim4500$ km s$^{-1}$) and a narrower (FWHM$\sim1400$ km s$^{-1}$) Gaussian component. The fit also includes a broad \ion{N}{V} doublet (F($\lambda1238.82$)/F($\lambda1242.80$) fixed to 2). Indeed, the emission feature at around 1270\AA\ (rest frame), rather than being a strongly redshifted \ion{Si}{II} blend, lies exactly at the expected red wing of a further disc line component from the \ion{N}{V} doublet, whose parameters are very similar to that of Ly$\alpha$, in particular the inclination angle and the inner radius (see Table~\ref{ngc3147_G140L_lines}).

\begin{figure}
	\includegraphics[width=\columnwidth]{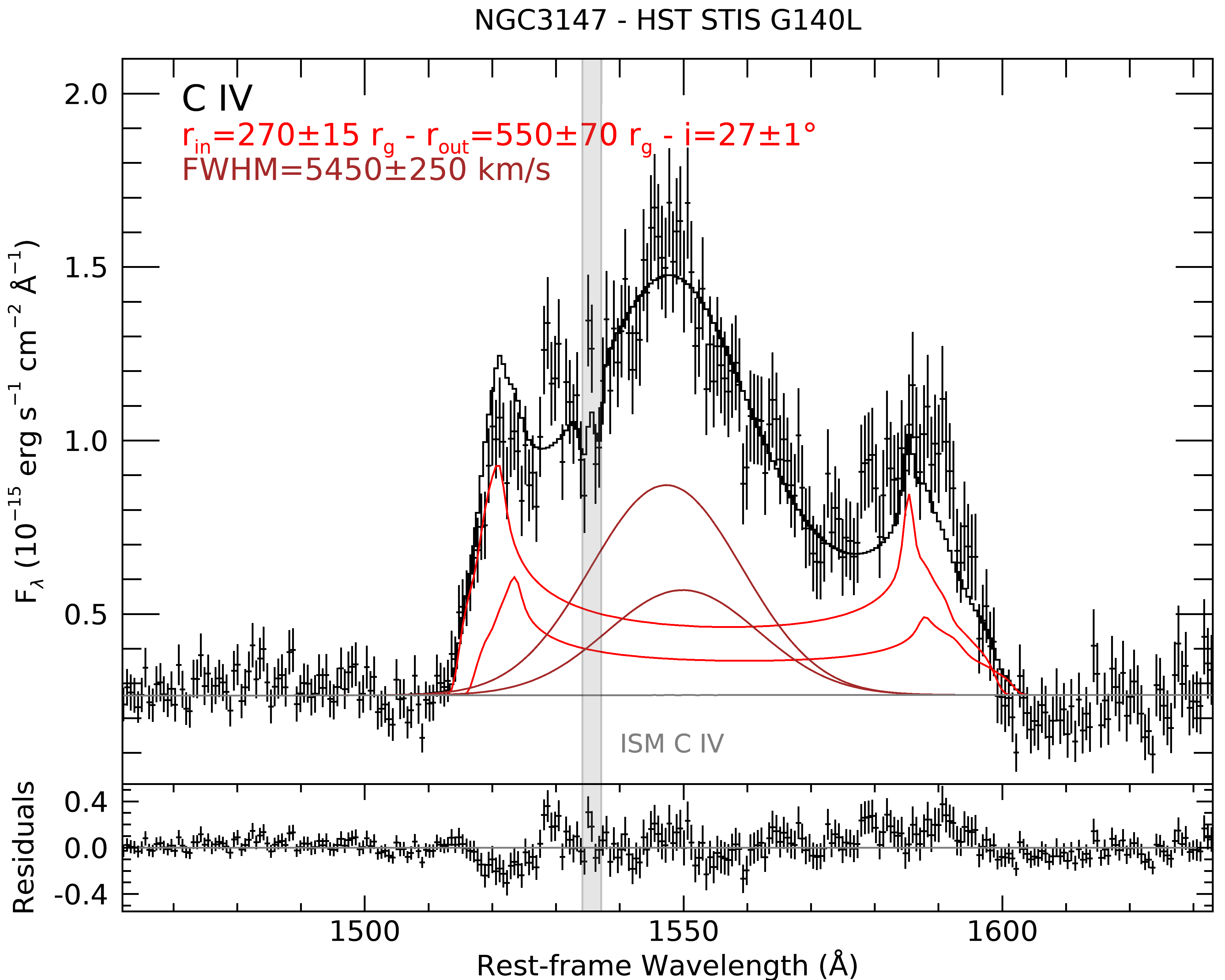}
    \caption{The \ion{C}{IV} emission line in the HST STIS G140L spectrum of NGC~3147. A detailed decomposition of the best-fitting model is shown, together with the main parameters (see text for details). The fit residuals are shown in the bottom panel. The foreground ISM absorption lines are highlighted by grey bands.}
    \label{fig:CIV}
\end{figure}

The \ion{C}{IV} emission line profile is also well modelled by a \textsc{Kerrdisk} disc line (see Fig.~\ref{fig:CIV}). Again, it is important to note how the inclination of $i=27\pm1\degr$ is also recovered in this case. The inner and outer radii of the disc are r$_{in}=270\pm15$ and r$_{out}=550\pm70$ r$_g$, as reported in Table~\ref{ngc3147_G140L_lines}. A further broad Gaussian line with an FWHM=$5450\pm250$ km s$^{-1}$ is also present, while a narrow component is not required by the fit. All the components are fitted with a doublet, fixing $F(\lambda1548.19)/F(\lambda1550.77)=2$.
Although the overall profile is well reproduced by this model, residuals in excess are still present at the red wing of the line profile. These can be modelled by adding a further Gaussian line at $\sim1586$ \AA\ (rest frame) with an FWHM$\sim2500$ km s$^{-1}$ and a flux of $\sim5.9\times10^{-15}$ erg cm$^{-2}$ s$^{-1}$. If interpreted as \ion{C}{IV} emission, the corresponding velocity shift would be $7100\pm200$ km s$^{-1}$.

\begin{figure}
	\includegraphics[width=\columnwidth]{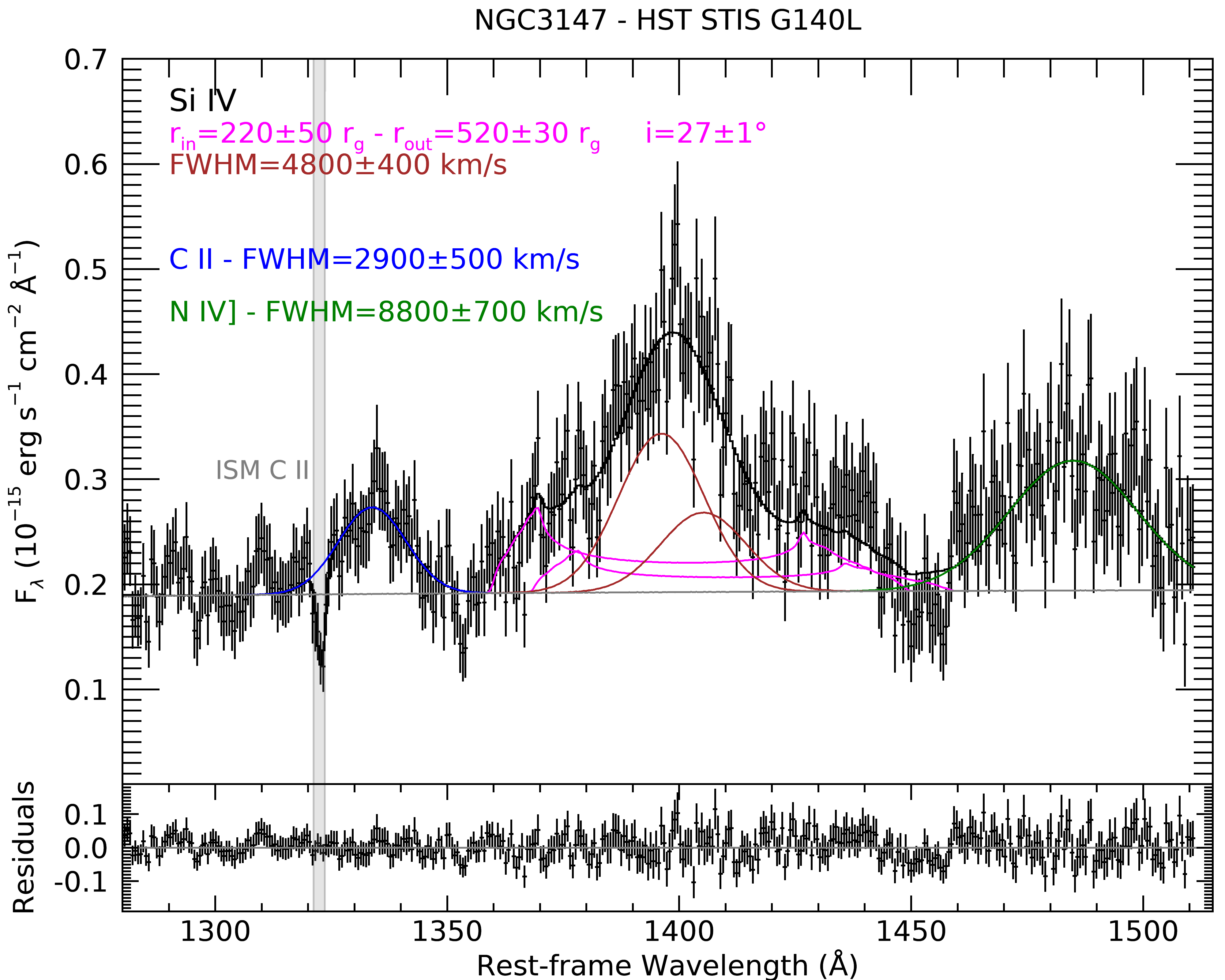}
    \caption{The \ion{Si}{IV} emission line in the HST STIS G140L spectrum of NGC~3147. A detailed decomposition of the best-fitting model is shown, together with the main parameters (see text for details). The fit residuals are shown in the bottom panel. The foreground ISM absorption lines are highlighted by grey bands.}
    \label{fig:SiIV}
\end{figure}

Finally, a disc line profile is a good modellization of the \ion{Si}{IV} doublet at 1393.76 and 1402.77 \AA\ (see Fig.~\ref{fig:SiIV}). Also in this case, the overall profile also includes a broad Gaussian component. A possible contribution from the \ion{O}{IV}] blend cannot be excluded, although not strictly required. The \ion{N}{IV}] line at 1486.5 \AA\ is also very broad, but the lower statistics and the vicinity to the huge \ion{C}{IV} line do not allow for a more complex profile decomposition.

\begin{figure*}
	\includegraphics[width=\columnwidth]{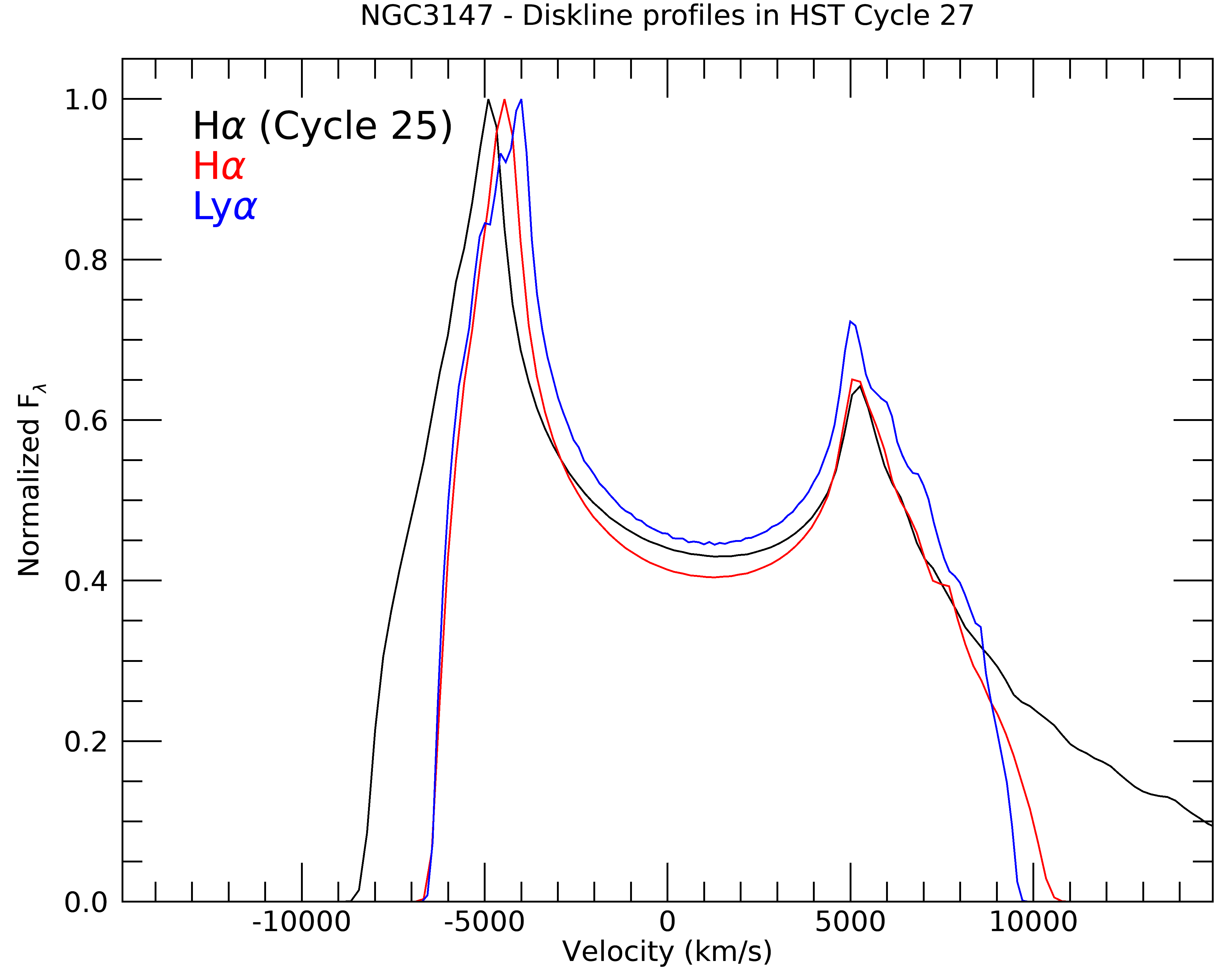}
	\includegraphics[width=\columnwidth]{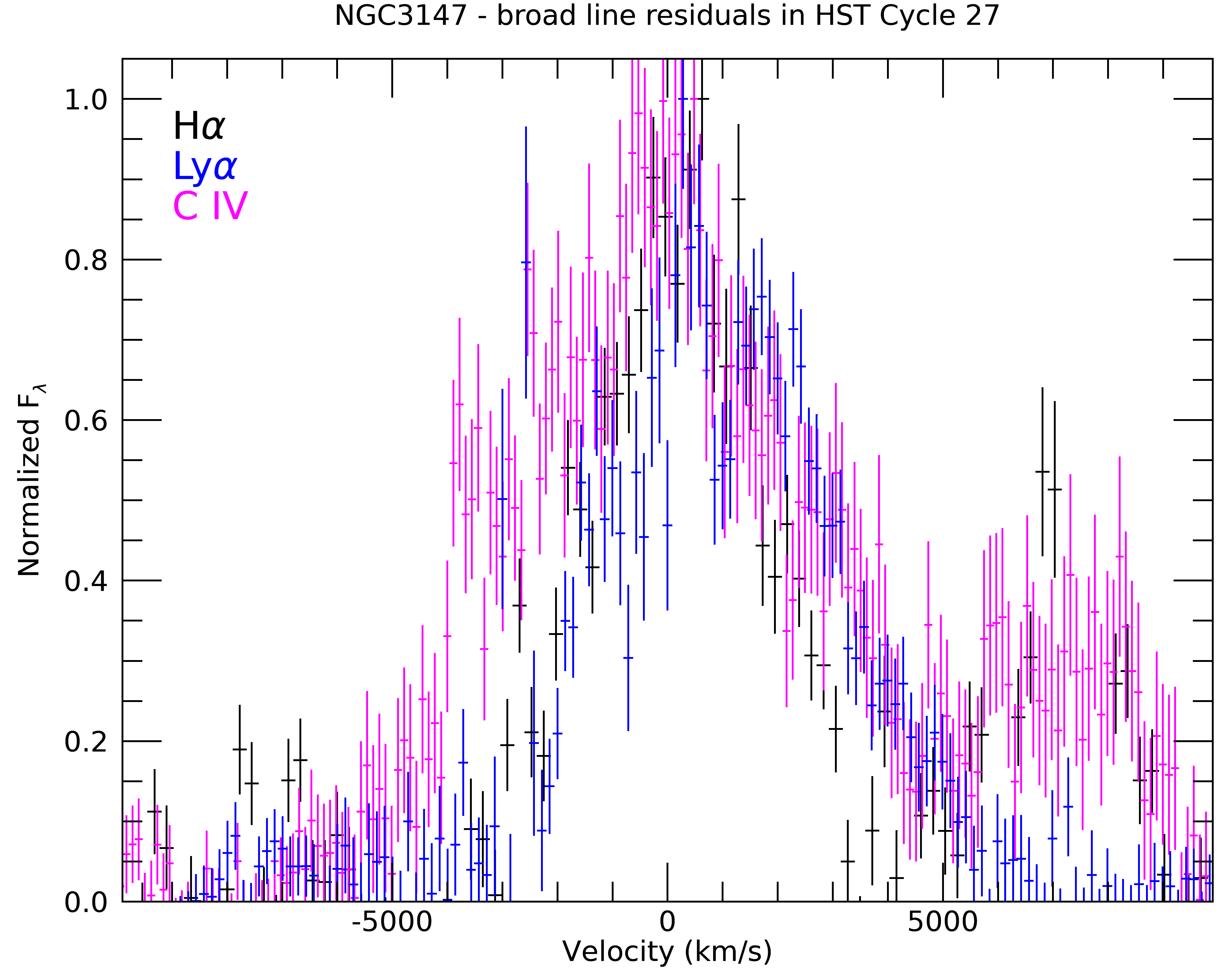}
    \caption{\textit{Left-hand panel:} velocity profile comparison between H$\alpha$ (STIS G750L, in black for the old epoch and red for the new epoch) and Ly$\alpha$ (STIS G140L, in blue) best-fitting disc line profiles, renormalized to their maximum flux. \textit{Right-hand panel}: velocity profile comparison between the \ion{C}{IV} (STIS G140L, in magenta), H$\alpha$ (STIS G750L, in black) and Ly$\alpha$ (STIS G140L, in blue) residuals, after taking into account the disc line component and the narrow emission lines from the NLR. Data are renormalized to their maximum flux.}
    \label{fig:CIV_Ha}
\end{figure*}

The disc line best-fitting profiles for H$\alpha$ and Ly$\alpha$ are directly compared in velocity space in Fig.~\ref{fig:CIV_Ha}, renormalized with respect to their maximum. From this plot, it is evident that the blueward extension of all the lines is in good agreement, as expected, being it mostly set by the inclination angle of the disc. Moreover, the red wing is also very similar for Ly$\alpha$ and H$\alpha$ taken at the same epoch, while note the much more extended red wing in the H$\alpha$ profile as observed in the previous epoch, due to the smaller inner radius. 

It is also interesting to look at the broad Gaussian components that are needed in all the modellizations of these emission lines. In Fig~\ref{fig:CIV_Ha} (right-hand panel), the residuals to the data after taking into account the disc line profiles and the narrow Gaussian components for the NLR are shown. The broad Gaussian component is apparent in all three emission lines. While this component appears quite consistent between H$\alpha$ and Ly$\alpha$, the residuals for \ion{C}{IV} are broader. However, it should be noted that the latter is a doublet, and this would also explain the apparent asymmetry of these residuals. Both H$\alpha$ and \ion{C}{IV} show a further component at $\sim7000$ km s$^{-1}$, as already noted in the previous fits. This component is not present in Ly$\alpha$.

\begin{table}
\caption{\label{ngc3147_G140L_lines} Emission line properties in the G140L spectrum. FWHMs and velocities are in km s$^{-1}$, radii in r$_g$ and dereddened fluxes in $10^{-15}$ erg cm$^{-2}$ s$^{-1}$.}
\begin{center}
\begin{tabular}{lllll}
\hline
Line & $\lambda_\mathrm{l}$ & FWHM & Flux & $v$ \\
\hline
Ly$\alpha$ & 1215.67 & $4500\pm150$ & $24.9\pm1.0$ & $780\pm80$\\
Ly$\alpha$ & 1215.67 & $1400\pm50$ & $17.6\pm0.6$ & $-120\pm30$\\
\multirow{2}{*}{\ion{N}{V}} &1238.82 & \multirow{2}{*}{$3750\pm250$} & $6.0\pm0.4$ & \multirow{2}{*}{$870\pm120$} \\
& 1242.80 &  & $3.0\pm0.2$ & \\
\ion{C}{II} & 1335.31 & $2900\pm500$ & $1.9\pm0.3$ & $130\pm120$\\
\multirow{2}{*}{\ion{Si}{IV}} & 1393.76 & \multirow{2}{*}{$4800\pm400$} & $4.2\pm0.4$ & \multirow{2}{*}{$660\pm200$}\\
& 1402.77 &  & $2.1\pm0.2$ & \\
\ion{N}{IV}] & 1486.5 & $8800\pm700$ & $5.2\pm0.6$ & $-160\pm350$\\
\multirow{2}{*}{\ion{C}{IV}} & 1548.19 & \multirow{2}{*}{$5450\pm250$} & $21.8\pm0.8$ & \multirow{2}{*}{$-170\pm120$}\\
& 1550.77 &  & $10.9\pm0.4$ & \\
\ion{He}{II} & 1640.5 & $3700\pm1700$ & $3.5\pm1.0$ & $0\pm600$\\
\ion{O}{III}] & 1664.14 & $3000\pm1000$ & $3.2\pm1.0$ & $300\pm800$\\
\hline
\multicolumn{5}{c}{\textsc{Broad disc line profiles}}\\
$\lambda_\mathrm{l}$ & $i$ & r$_{\rm in}$ & r$_{\rm out}$ & Flux \\
\hline
\multicolumn{5}{c}{\textsc{Ly$\alpha$}}\\
1215.67 & $27\pm1$& $300\pm5$ & $1020\pm50$ & $33.5\pm1.4$ \\
\hline
\multicolumn{5}{c}{\textsc{\ion{N}{V}}}\\
1238.82 & \multirow{2}{*}{$27\pm1$}& \multirow{2}{*}{$305\pm5$} & \multirow{2}{*}{$460\pm20$} &  $7.0\pm0.4$\\
1242.80 & & & & $3.5\pm0.2$\\
\hline
\multicolumn{5}{c}{\textsc{\ion{Si}{IV}}}\\
1393.76 & \multirow{2}{*}{$27\pm1$}& \multirow{2}{*}{$220\pm50$} & \multirow{2}{*}{$520\pm30$} &  $3.7\pm0.4$\\
1402.77 & & & & $1.9\pm0.2$\\
\hline
\multicolumn{5}{c}{\textsc{\ion{C}{IV}}}\\
1548.19 & \multirow{2}{*}{$27\pm1$}& \multirow{2}{*}{$270\pm15$} & \multirow{2}{*}{$550\pm70$} &  $26.4\pm1.0$\\
1550.77 & & & & $13.2\pm0.5$\\
\hline

 
\end{tabular}
\end{center}
\end{table}

\subsection{\label{sed}The nuclear SED}

\begin{figure*}
    \includegraphics[width=\columnwidth]{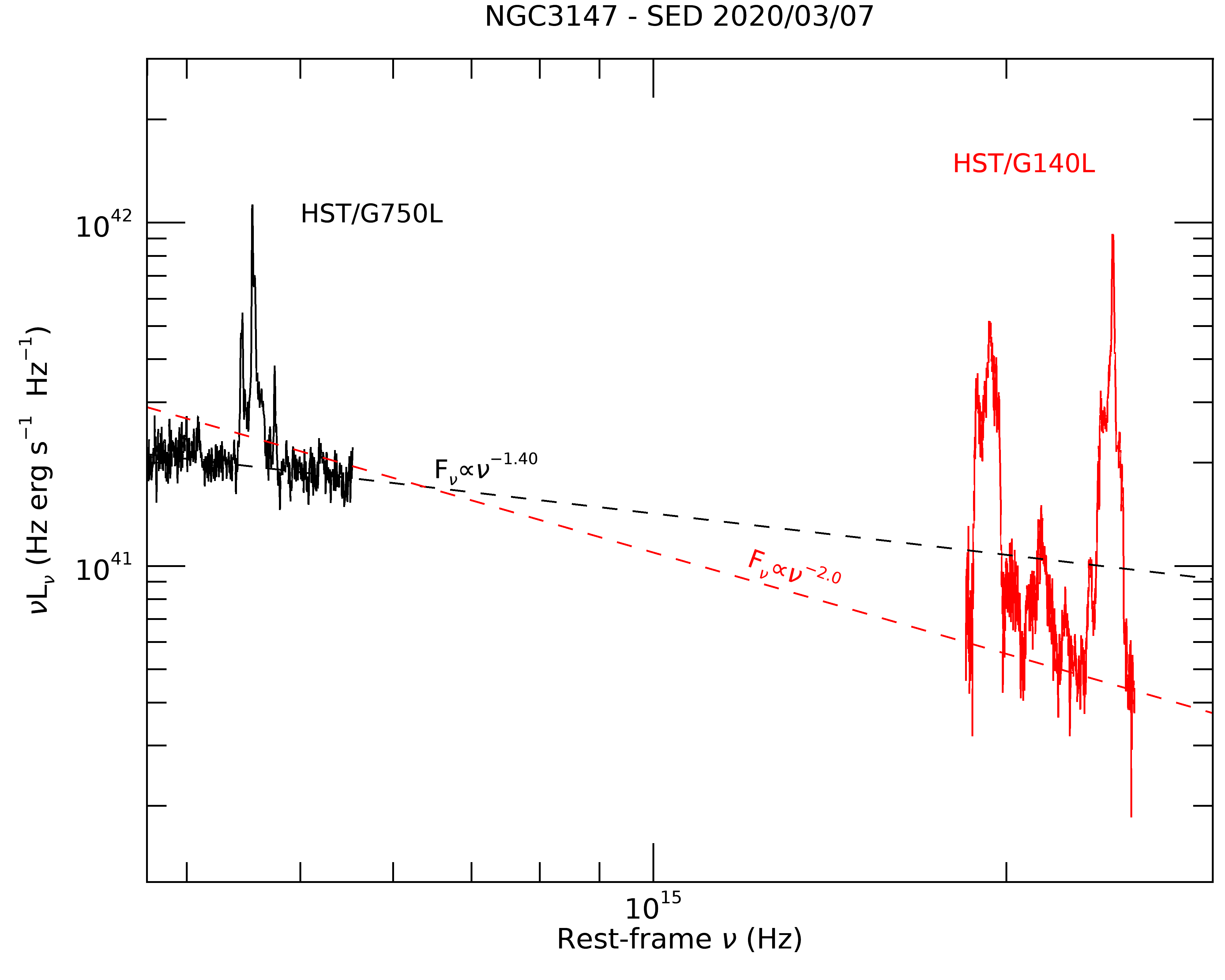} \includegraphics[width=\columnwidth]{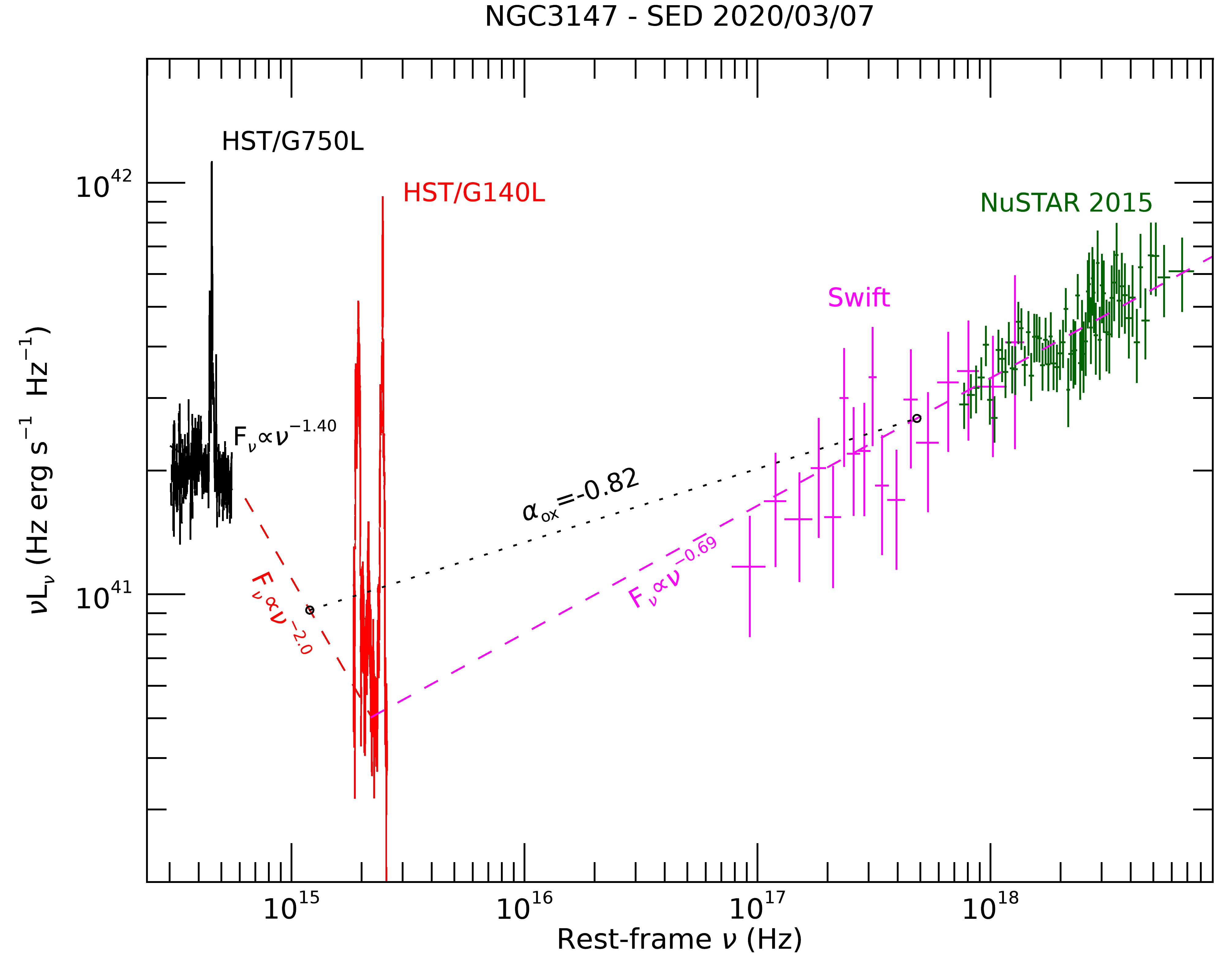}
    \caption{The observed SED of NGC~3147 with the simultaneous HST and Swift observations in 2020 March. Data are rebinned for visual purposes, and the geocoronal Ly$\alpha$ emission line is removed from the G140L spectrum. The best-fitting power law for the continuum of the HST spectra is overimposed in blue, while the $\alpha_{ox}$ in black. The NuSTAR data are instead taken in 2015, while the MIR data from Gemini/Michelle in 2010.}
    \label{fig:broadband}
\end{figure*}

Thanks to the very small slit adopted for the \textit{HST} observation, we can build the first nuclear SED of NGC~3147. The continua observed in the \textit{HST} G750L and G140L are well described by power laws with indices (in $F_\nu$) of $-1.40\pm0.04$ and $-2.0^{+0.3}_{-0.4}$, respectively, as derived in Sect.~\ref{optan} and \ref{uvan} (see the left-hand panel of Fig.~\ref{fig:broadband}). On the other hand, the simultaneous \textit{Swift} X-ray observation is well described by a power law with the same photon index measured in the 2015 NuSTAR observation \citep[$\Gamma=1.69\pm0.05$,][]{Bianchi2017a}. The 2-10 keV flux is $2.5\pm0.2\times10^{-12}$ erg cm$^{-2}$ s$^{-1}$, i.e. at the same level of the NuSTAR observation and just above the historical average flux of the source \citep{Bianchi2017a}.

The optical-to-X-ray SED is shown in the right-hand panel of Fig.~\ref{fig:broadband}. It is evident that it is very hard, with a peak (in $\nu L_\nu$) in the hard X-rays. By combining the \textit{Swift} 2 keV observed density flux with the 2500\AA\ density flux as extrapolated from the G140L spectrum, we obtain $\alpha_{ox}=-0.82\pm0.16$ (see the right-hand panel of Fig.~\ref{fig:broadband}). This value is significantly flatter than the one derived by \citet{Bianchi2008c} from the XMM-\textit{Newton} Optical Monitor, which is dominated by the emission from the host galaxy, and is therefore not representative of the nuclear emission. For the same reason, the new \textit{Swift} Ultra-violet Optical Telescope (UVOT) data are not displayed here, and cannot be used to build the nuclear SED.

Finally, we note that in the literature there are subarcsec near-infrared (NIR) and mid-infrared (MIR) fluxes of NGC~3147 taken with Gemini/Michelle in 2010 at 11.6 and 12.5 $\mu$m \citep[][]{Asmus2014,Mason2015}, which would show a rise of the SED with respect to the optical emission. The resulting IR-to-optical power law would have an index (in $F_\nu$) of -1.66. However, the IR emission in NGC~3147 appears to be host-dominated also at subarcsec resolution, hence we decided not to include these data in the SED.

\section{Discussion}

\subsection{The disc and the BLR}

The presence of a thin accretion disc in the low-accreting AGN NGC~3147 was inferred by \citet{Bianchi2019a} with the discovery of a broad H$\alpha$ emission line with a disc line profile. An H$\alpha$ line with the same profile is confirmed in the \textit{HST} spectra presented in this work, taken 2 yr after, suggesting that the disc is persistent. Moreover, broad disc line profiles are observed for other prominent emission lines in UV, namely Ly$\alpha$, \ion{C}{iv}, \ion{N}{v} and \ion{Si}{IV}. The thin disc interpretation is strongly reinforced by the fact that all these profiles present blue and red peak locations well matched with simple disc line predictions. Moreover, they all agree with a disc inclination between 23\degr and 27\degr.

The H$\alpha$ disc line component was much stronger in Cycle 25, although this does not appear to be accompanied by a significant flux variation of the source. The disc line variability may be due to a changing structure in the BLR, expected in a dynamical time-scale that could be of the order of $\sim1$ month in such a compact BLR. A sort of `accretion event' may have occurred in Cycle 25, with gas swirling in at $\sim100$ r$_g$, and now we only see residual material further out. This scenario is reinforced by the larger inner radius measured in the new observation, indeed suggesting that the disc receded from $\sim80$ to $\sim200$ r$_g$. The disc line profiles for the UV lines are in good agreement with H$\alpha$ (notably for the inclination angle), and are all well modelled by inner radii between 200 and 300 r$_g$.

All the broad-line profiles require a Gaussian component in addition to the disc line, with a very similar FWHM around $4000-5000$ km s$^{-1}$. The gas that produces these lines should be at around $4000-5000$ r$_g$ (neglecting any virial factor), i.e. $\sim0.06-0.07$ pc for $\log\mathrm{M_{BH}/M_\odot}=8.49$. Since the sublimation radius for NGC~3147 should be located not farther than $\sim0.03$ pc \citep[][but considering the whole bolometric luminosity here]{Barvainis1987a}, this means that dust is expected to be present, and this broad component cannot be associated with a classical BLR. However, it may be possible that the very hard SED that characterizes this source is more efficient at destroying dust at farther distances.

\subsection{The NLR}

The stellar velocity dispersion in NGC~3147 is $\sigma_*=233$ km s$^{-1}$ \citep{VanDenBosch2015}, so we expect the FWHM of lines produced outside the sphere of influence of the central BH to be of the order of $2.35\sigma_*\sim550$ km s$^{-1}$. This is indeed what is measured for [\ion{N}{II}], H$\alpha$ and [\ion{S}{II}]. On the other hand, [\ion{O}{I}] is significantly broader. This is qualitatively in agreement with a correlation with the critical density, as already shown in \cite{Filippenko1984}, since [\ion{O}{I}] has the largest critical density and is therefore expected to be produced in a denser gas, closer to the BH. Indeed, a natural consequence of a radiation pressure compressed (RPC) NLR is that the lines with the highest critical density should be produced inside the BH sphere of influence, when the luminosity is low enough with respect to the BH mass \citep{Stern2014b}. Using $\log\mathrm{M_{BH}/M_\odot}=8.49$, $L_\mathrm{ion}=L_\mathrm{bol}=2.9\times10^{42}$ erg s$^{-1}$ (see Sect.~\ref{sec:sed}), and a critical density $1.1\times10^6$ cm$^{-3}$ \citep[][]{Martin2015}, we can estimate FWHM$_{[\ion{O}{I}]}=1400$ km s$^{-1}$ \citep[eq. 28 in][]{Stern2014b}. Note that this is the maximum expected FWHM if all [\ion{O}{I}] emission is produced at the critical density. In our spectrum, it appears that the bulk of the line is produced at farther distances, where the emissivity is still very high \citep[see Fig.~6 in][]{Stern2014b}. However, [\ion{O}{I}] can potentially have a broader base. 

It is interesting to check if the narrow lines' ratios derived from our small-aperture \textit{HST} spectrum are consistent with a Seyfert-like nucleus. From Table~\ref{ngc3147_lines} we get $\log$ [\ion{N}{II}]/H$\alpha=0.528\pm0.012$, which is in rough agreement with what found in larger aperture, ground-based spectra \citep[0.3-0.43:][]{Ho1997,Bianchi2012}, and well inside the AGN region in standard BPT diagrams \citep[e.g.][]{Kewley2006}. On the other hand, we have $\log$ [\ion{S}{II}]/H$\alpha=0.20\pm0.02$ and $\log$ [\ion{O}{I}]/H$\alpha=-0.10\pm0.02$, which are larger than 0.06 \citep[but it is 0.2 in ][]{Bianchi2012} and -0.82, respectively, as found in \citet{Ho1997}. These values are more suggestive of a LINER classification with respect to a Seyfert, unless $\log$[\ion{O}{III}]/H$\beta$ is significantly larger than 0.79-0.85, as measured in \citet{Ho1997} and \citet{Bianchi2012}. Unfortunately, our \textit{HST} spectra do not cover the [\ion{O}{III}] and H$\beta$ band, so we cannot give the final word on the matter. However, we note that the electron density derived from the [\ion{S}{II}] doublet, i.e. $n_e=680\pm80$ cm$^{-3}$, is significantly larger than what is found on average in LINERs, and more in agreement with the typical densities in Seyferts \citep[e.g.][]{Ho2008}.

\subsection{\label{sec:sed}The SED}

\begin{figure}
    \includegraphics[width=\columnwidth]{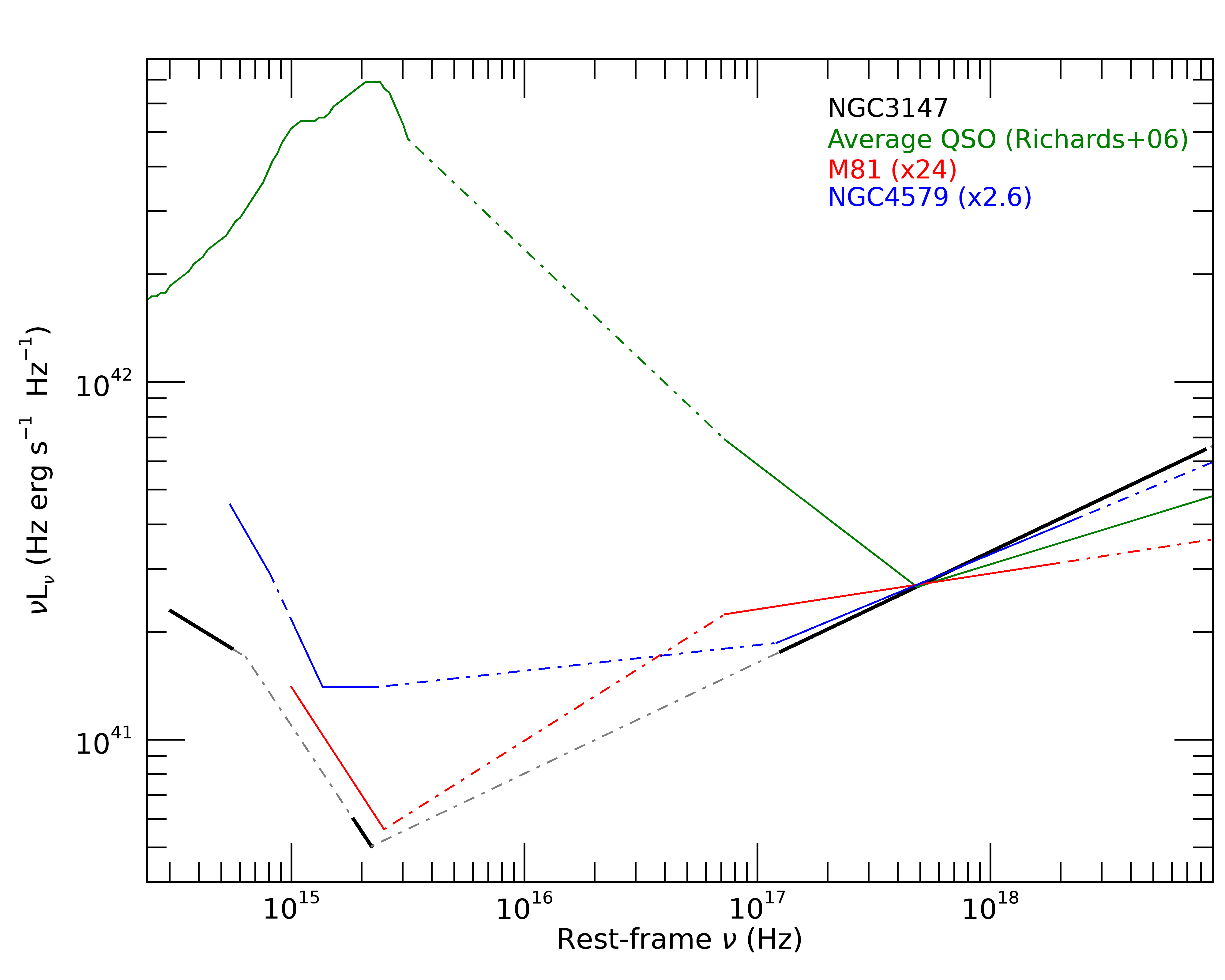}
    \caption{The SED of NGC~3147 (black) as parametrized in Fig.~\ref{fig:broadband} is plotted with the standard average SED of Type 1 Quasars \citep{Richards2006} in green. The SEDs of two representative LLAGN (with double-peaked lines and \textit{HST} optical/UV spectroscopic observations) are also shown with different colours (see the legend). All SEDs are re-normalized at the 2 keV luminosity density of NGC~3147 (the corresponding renormalization factor is shown in the legend), the solid lines are used in the frequency ranges where there are data, while the extrapolations are dot-dashed. Note that, differently from NGC~3147, the data in the other SEDs are not simultaneous. See text for details.}
    \label{fig:SED_compare}
\end{figure}

We have reconstructed the nuclear SED of NGC~3147 in Sect.~\ref{sed}. Notably, the optical, UV and X-ray data are taken simultaneously, and the \textit{HST} data have subarcsec resolution, minimizing the host contamination\footnote{Although the \textit{Swift} and \textit{NuSTAR} have much larger spatial resolution, \textit{Chandra} data and rapid variability conclusively showed that the X-ray emission in NGC~3147 is dominated by the nucleus \citep[][]{Bianchi2017a}}. The resulting SED, as parametrized with power laws in different bands, is shown again in Fig.~\ref{fig:SED_compare}: a spectral index of -1.40 for the optical band, a steepening to -2.0 in the UV and a rise to -0.69 up to the hard X-rays. An integral of this SED from 1 $\mu$m to 200 keV gives a bolometric luminosity of $2.9\times10^{42}$ erg s$^{-1}$, where we have extrapolated the X-ray power law introducing an exponential cutoff at 150 keV, as found on average in local Seyfert galaxies \citep[e.g.][]{Tortosa2018,Middei2019}. With this revised bolometric luminosity, the Eddington ratio of NGC~3147 is  $L_\mathrm{bol}/L_\mathrm{Edd}\sim7.5\times10^{-5}$, at the lower end of previous estimated ranges \citep[e.g.][]{Bianchi2019a}. Considering the 2-10 keV luminosity observed in the \textit{Swift} observation ($5.6\times10^{41}$ erg s$^{-1}$), this gives an X-ray bolometric correction of $\sim5$, which is much lower than standard values, even at the low-luminosity end of the AGN population \citep[see e.g.][]{Duras2017}. Most of the bolometric luminosity in NGC~3147 is released in the X-rays. Therefore, although the presence of double-peaked optical and UV lines shows that there is a thin disc in this source, this appears to be a rather `passive' one, and most of the radiation is instead produced in some hot medium.

Indeed, the most striking characteristic of this SED is the lack of a UV bump. This is evident when compared to the standard SED of AGN \citep[as in][]{Richards2006}, shown in green in the same figure, after renormalizing it at 2 keV\footnote{Above $\log \nu=15.5$ this SED is extrapolated in order to have an $\alpha_{ox}=-1.5$, a $-1.5$ index to model the `soft excess' between 0.3 and 2 keV, then an index of $-0.8$ is adopted above 2 keV.}. Indeed, in NGC~3147 the minimum energy is released in the UV, contrary to what found for a normal AGN. This reflects in a much flatter $\alpha_{ox}$ (-0.82) with respect to standard AGN \citep[roughly between -1.2 and -1.6: e.g.][]{Grupe2010,Vagnetti2010}. 

Such flat $\alpha_{ox}$ are instead more common in LLAGN \citep[e.g.][]{Ho1996,Ho2000c,Sabra2003,Maoz2007,Ho2008,Ontiveros22}. Indeed, the $\alpha_{ox}$ of luminous AGN is known to follow a tight inverse relation with luminosity \citep[see e.g.][]{Just2007a}. Extrapolating the relation found in \citet{Just2007a} to the low UV luminosity of NGC~3147 (L$_{2500\text{\AA}}\sim7.6\times10^{25}$ erg s$^{-1}$ Hz$^{-1}$ as extrapolated from our \textit{HST} UV spectrum), we get $\alpha_{ox}=-0.92$, which is very close to what we actually observe in this source. It is quite impressive that the $\alpha_{ox}$ luminosity relation appears to extend down to these bolometric luminosities. This suggests that there is no clear transition from luminous AGN to LLAGN, but just a progressive systematic trend in the relative contribution of the X-rays to the UV bump, with the latter becoming weaker and colder at decreasing luminosity.

We can derive the frequency where a standard \citet{Shakura1973} disc should peak in the case of NGC~3147. From the bolometric luminosity derived above, we can estimate the mass accretion rate assuming efficiencies $\eta=0.057$ and $0.321$, i.e. as expected if the inner radius goes down to the last stable orbit, for $a=0$ and $a=0.998$, respectively \citep[][]{NovikovThorne1973}. We get $\dot{\mathrm{M}}=1.1\times10^{-3}$ and $2.0\times10^{-4}$  M$_\odot$/yr\footnote{We drop the factor 2, valid only in the Newtonian approximation \citep[e.g.][]{Davis2011}, and assume an inclination $\cos{i}=0.8$.}. These values can be used to derive the maximum disc temperature, as in \citet{Laor2011a}, and the relative frequency peak of the emission (in $\nu L_\nu$):  $4.1\times10^{14}$ and $6.4\times10^{14}$ Hz, again for a non-rotating and a maximally rotating BH. This frequency range ($0.47-0.74 \mu$m) corresponds approximately to the band covered by our \textit{HST} optical spectra. Indeed, the flat (-1) slope we observe in the optical may be the accretion disc emission close to the peak. It is then possible that the observed steepening at higher frequencies to the UV is indeed the SED on the blue side of the accretion disc peak. In normal AGN this blue side is shortwards of $\sim1000$\AA, and is also a steep power law \citep[e.g.][]{Laor2014}. 

Therefore, the steep -2 UV power law we observe in NGC~3147 may be Comptonization of the cold disc by a warm corona, in analogy to one of the standard interpretations of the so-called `soft excess' observed in the X-ray emission of more radiatively efficient AGN \citep[e.g.][]{Petrucci2018,Petrucci2020}. In NGC~3147, this component is expected to shift to lower frequencies, together with the accretion disc peak, and indeed there is no clear evidence for an X-ray soft excess in NGC~3147 \citep{Bianchi2012,Bianchi2017a}, although there are some hints of a different variability between the soft and the hard X-ray bands \citep[][]{Matt2012}. 

We searched in the literature for other LLAGN with double-peaked optical/UV emission lines and subarcsec \textit{HST} UV observations allowing for a direct estimate of the intrinsic continuum slope in this band. We found two sources satisfying these criteria: M81 and NGC4579. Their SEDs were reconstructed using the optical/UV slopes derived from \textit{HST} in \citet{Ho1996} (M81), \citet{Barth1996} and \citet{Molina2018a} (NGC4579), the X-ray slopes observed by XMM-\textit{Newton} for M81 \citep[][]{Cappi2006} and NGC4579 \citep[][]{Bianchi2009}, and the $a_{ox}$ reported by \citet{Ho1996} (M81) and \citet{Barth1996} (NGC4579). The resulting SEDs are shown in Fig.~\ref{fig:SED_compare}. The overall shape of the SEDs of LLAGN is very similar to that of NGC~3147: a dominant X-ray emission and a weak UV emission characterized by a steep power law. 

The inclusion of IR data to the SEDs of LLAGN generally leads to a rise of the emission up to around 10 $\mu$m \citep[see e.g.][]{Ho2008,Ontiveros22}. While we are not confident that the existing subarcsec MIR and NIR photometry for NGC~3147 is dominated by the AGN, such a rise is observed also in NGC~3147, as briefly mentioned in Sect.~\ref{sed}. The IR emission in luminous AGN is dominated by dust reprocessing, generally associated to the torus. Indeed, NGC~3147 has a strong iron K$\alpha$ line in its X-ray spectrum \citep[][]{Bianchi2012,Bianchi2017a,Matt2012}, which suggests a significant contribution from Compton-thick neutral circumnuclear matter, as routinely found for more luminous AGN \citep[e.g.][]{Bianchi2009}. However, hot dust is mostly heated by the optical-UV emission, so it is unlikely that the IR luminosity peak is much stronger than the UV peak. On the other hand, the IR emission in NGC~3147 can be significantly contributed by the accretion disc emission below the peak. This could be tested with future observations with NIR adaptive optics, as done in other very low luminosity AGN \citep{Dumont2020}.

\section{Conclusions}

We have reported on a new \textit{HST} observation of NGC~3147, a low-luminosity ($\mathrm{L_{bol}}\sim10^{42}$ erg s$^{-1}$), low-accreting ($\mathrm{L_{bol}/L_{Edd}}\sim10^{-4}$) AGN, once considered the archetypal true type Seyfert 2. The main results can be summarized as follows.

\begin{itemize}
\item The broad, double-peaked, disc line H$\alpha$ profile revealed by \citet{Bianchi2019a} is confirmed, further showing variability both in flux and in the inner radius of the emitting disc with respect to the previous observation.
\item disc line profiles are found for the first time also in prominent UV lines, in particular Ly$\alpha$ and \ion{C}{iv}, with disc parameters in agreement with those derived for H$\alpha$, in particular the inclination angle, constrained to be around 23-27\degr.
\item The simultaneous subarcsec optical-to-X-ray SED is characterized by the absence of a thermal UV bump, and an emission peak in the X-rays. 
\end{itemize}

The overall scenario seems to indicate that NGC~3147 is a prototypical LLAGN, characterized by a weaker and colder thin accretion disc with respect to more luminous AGN. Most of its extreme properties can be indeed derived from simple extrapolations of well-known standard relations to a regime characterized by low luminosity and accretion rate.

\section*{Acknowledgements}
We would like to thank the referee, Jules Halpern, for a constructive review that improved the paper, and Juan Antonio Fern\'andez Ontiveros and Michal Zaja\v{c}ek for very useful and stimulating discussions.
SB acknowledges financial support from the Italian Space Agency under the grant ASI-INAF 2017-14-H.O. RA and AB would like to thank Michal Zaja\v{c}ek for very useful discussions. This research is based on observations made with the NASA/ESA Hubble Space Telescope obtained from the Space Telescope Science Institute, which is operated by the Association of Universities for Research in Astronomy, Inc., under NASA contract NAS 5–26555. 
We thank the \textit{Swift} team for the Target of Opportunity observation. Part of this work is based on archival data, software or online services provided by the Space Science Data Center - ASI.


\section*{Data Availability}

These observations are associated with programmes GO 15225 and GO 15908. The \textit{HST} data described here may be obtained from the MAST archive at \url{http://dx.doi.org/10.17909/t9-7vzv-2j06} and \url{http://dx.doi.org/10.17909/nwrd-9n20}.



\bibliographystyle{mnras}
\bibliography{ngc3147_new} 




\appendix

\section{The AGN continuum contribution to the total red continuum.}
\label{host}
The STIS reduction pipeline and appropriate flux calibration are complex procedures, since the corrections depend on the nature of the target. The STIS pipeline is best suited to either an isolated point source, or a relatively homogeneous extended target within the slit. For the former, the pipeline is optimized for a 7-pixel extraction height, instead of the adopted 3-pixel aperture, as described in Sect.~\ref{datared}. This is necessary because the image rectification and resampling in the STIS reduction pipeline smooths the data substantially in the spatial direction\footnote{On page 179 of the Data Handbook, we read that "In general we note that the cross dispersion profiles can be quite extended (particularly in the far-UV and in the near-IR.)" } \citep[Section 5.4 of][]{Sohn2019}. However, this process precludes a direct way to estimate any host contamination to the point source.

Thus we abandoned decomposition of the point source from the host using the slit spectra, and rely exclusively on our 120s 2D image of our target in comparison with a stellar image from another programme. The star was G158-100, observed with the same optical elements (Proposal ID: 7642).
We then add some corroborative arguments that give evidence that the red continuum in the G750L spectrum is in fact starlight-dominated. Note that we also specifically looked for late-type stellar absorption features such as the \ion{Na}{I} D doublet and the \ion{Ca}{ii} triplet in the G750L spectrum, but they are not detected, and we were not able to place useful constraints in this way.
We present instead three methods as follows: 
\subsection{Image decomposition using a stellar PSF}
\begin{figure}
	\includegraphics[width=\columnwidth]{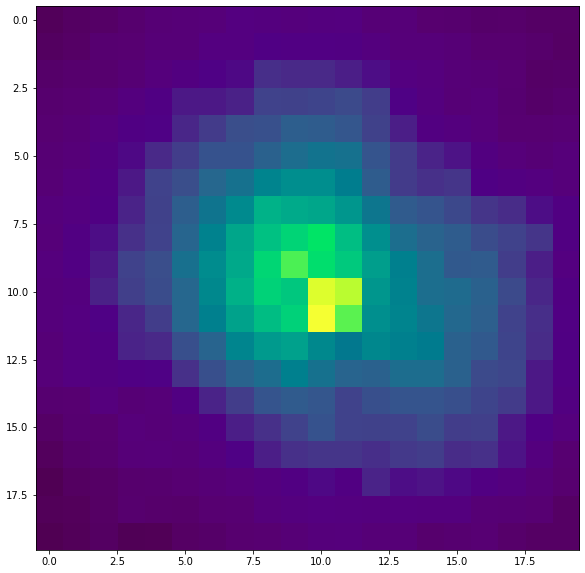}
    \caption{The residual flat top of NGC3147 which was achieved by a series of PSF subtractions (of G158-100) on the NGC3147 image. Both the images i.e NGC3147 and the star G158-100 were taken with same optical elements and detector}
    \label{fig:residual}
\end{figure}

We experimentally subtracted the 2D PSF profiles from the NGC~3147 image with various normalizations until the conspicuous spike in the NGC~3147 image was gone, and the residual was roughly flat-topped (Fig.\ref{fig:residual}). That particular normalization of the PSF provides an approximate upper limit to the nuclear contribution. Our value for this upper limit to the fractional contribution of an AGN  point source in the displayed $0.1\arcsec \times 0.15\arcsec$ spectrum shown in Fig.\ref{fig:kerrdisk} is $0.30\pm 0.05$. It is an upper limit in part because real galaxies are peaked in the center to varying degrees. Also, the broad H$\alpha$ line equivalent width is potentially significant, and strong, compact, and narrow lines are present. They all contribute to the fitted point source, along with the continuum. We ameliorated this somewhat by using the Epoch 2 data when the broad line was weaker. The broad-line equivalent width was approximately 449{\AA}$\pm 13${\AA} in Epoch 1 \citep{Bianchi2019a}. We can only estimate the effective bandpass of this long-pass filter, considering its transmission, the detector efficiency, and the SED. Significant light contribution comes from all the way from 5200{\AA} to 10200{\AA}, but the S/N and thus the light contribution decreases strongly past$\sim 7500${\AA}. Given the ratio of the summed line equivalent widths to the continuum range observed with high sensitivity, we conclude that the lines contribute only $\sim 20\pm 5\%$ of the AGN point source. Thus we adopt a fraction of $(0.30\pm 0.05)\times (0.80\pm 0.05) =  24\pm 5\%$  for the AGN continuum contribution to our total G750L continuum in the aperture.
\subsection{Continuum and H$\alpha$ variability}
The flux of the broad H$\alpha$ line component we attribute to emission from a thin disc as noted in Sec. \ref{optan} has decreased by about a factor of 3 over the 660 d between our epochs.  Based on vast experience at higher Eddington ratios, as well as physical plausibility, one can posit that the nuclear continuum around the Balmer lines changed by a factor similar to or greater than that of the line \citep{2021MNRAS.508.6077G}.  We find that the total red continuum has not changed significantly, with an upper limit to any decrease of about 10\%.

Suppose the AGN continuum decreased by anything like the factor of 3 decrease in the disc line, and that it drives that line. Since the total continuum declined by 10\% or less, it must have contributed no more than around 30\% of the continuum in Epoch 1. Note though that this does not do justice to the complexity of the situation because the Gaussian component of the broad H$\alpha$ line rose instead in flux. Nevertheless in this situation it seems unlikely that the AGN continuum is constant, suggesting significant dilution from a constant component, i.e. starlight.

\subsection{Extrapolation of starlight from the inner arcsec to the small HST subaperture.}
The starlight in the aperture of ground-based spectra strongly constrains the amount of starlight in the \textit{HST} aperture, based on the known photometric properties of galaxies. We note that the superb S/N total flux spectrum accumulated by \citet{Tran2011} in measuring polarization is nearly pure starlight. Their continuum flux is around fifteen times ours, thus at least 93\% of their light is outside our aperture and almost certainly starlight. Moreover, some of the light common to both apertures is starlight.  Also recall that there is no trace of the broad line in their data, but just strong stellar absorption lines, supporting nearly pure host light in their spectrum.  Thus their starlight fraction is very close to 100\%.

Based on the aperture solid angles, and the very similar ratios of the fluxes, host starlight would have to be important in the \textit{HST} aperture even if the host surface brightness were uniform and not concentrated to the center at all.
Based on actual surface photometry of galaxies, it must be that the starlight in our aperture is similar to our observed continuum. 

A high-quality study of the inner regions of the hosts of a sample of LLAGN \citep{2013ApJ...767..149B} found that the enclosed starlight as a function of metric aperture in the relevant region follows approximately $L_{enclosed} \propto R$. Unless our host is unique, that means our aperture ($0.1\arcsec \times 0.15\arcsec$) compared to Tran's ($1\arcsec \times 1\arcsec$) means our starlight is lower than theirs by a factor of 10\footnote{The slit extraction height is not reported explicitly in \citet{Tran2011}. However, even assuming a much larger height, as the 2.7\arcsec\ reported in \citet{Tran2010a} for NGC~2110 for a similar Keck configuration, this value only raises to $\sim20$.}.  Since our continuum is lower by a factor of 15, we conclude again that starlight likely dominates.



\bsp	
\label{lastpage}
\end{document}